\documentclass{elsart}

\usepackage{graphicx}
\usepackage{epsfig}

\usepackage{amssymb}

\begin{document}

\begin{frontmatter}

\title{Thermal Bremsstrahlung photons probing the nuclear caloric curve}

\author[SUBATECH]{D.G.~d'Enterria\thanksref{UAB}},
\ead{enterria@in2p3.fr}
\author[SUBATECH]{G.~Mart\'{\i}nez},
\author[SUBATECH]{L.~Aphecetche},
\author[SUBATECH]{H.~Delagrange},
\author[UAB]{F.~Fern\'andez},
\author[KVI]{H.~L\"ohner},
\author[UAB]{R.~Ortega},
\author[KVI]{R.~Ostendorf},
\author[SUBATECH]{Y.~Schutz},
\author[KVI]{H.W.~Wilschut}

\address[SUBATECH]{SUBATECH, 4 rue A. Kastler, 44307 Nantes, France}
\address[KVI]{Kernfysisch Versneller Instituut, 9747 AA Groningen, The Netherlands}
\address[UAB]{Grup de F\'{\i}sica de les Radiacions, Universitat Aut\`onoma de Barcelona,
 08193 Cerdanyola del Vall\`es, Catalonia}






\begin{abstract}
Hard-photon (E$_{\gamma}>$ 30 MeV) emission from second-chance 
nucleon-nucleon Bremsstrahlung collisions in intermediate energy 
heavy-ion reactions is studied employing a realistic thermal model. 
Photon spectra and yields measured in several nucleus-nucleus reactions
are consistent with an emission from hot nuclear systems
with temperatures $T\approx$ 4 - 7 MeV. The corresponding caloric curve
in the region of excitation energies $\epsilon^\star\approx$ 3{\it A} 
- 8{\it A} MeV shows lower values of $T$ than those expected for a
Fermi fluid.
\end{abstract}

\end{frontmatter}

\section{Introduction}
The determination of the thermodynamical properties such as temperature,
density, and excitation energy of the hot nuclear systems
produced in nucleus-nucleus reactions is one of the main goals of
heavy-ion (HI) physics. At moderate excitation energies, $\epsilon^\star\approx$
3{\it A} - 15{\it A} MeV, the experimental derivation of these
observables as well as their correlation is a prerequisite for a
quantitative investigation of the nuclear equation of state $\epsilon$ =
$\epsilon$($\rho$,$T$) in connection with a possible liquid-gas
phase transition \cite{enterria:Poch97,enterria:Gupt00,enterria:Rich00,enterria:Hirs99}. 
To date the most unambiguous evidence for such
a phase transition in HI collisions is given by the ``caloric curve'' 
\cite{enterria:Poch97} 
which relates the thermal energy of an excited
nucleus to its temperature, $\epsilon^\star$ = $\epsilon^\star$($T$).
For very low $\epsilon^\star$, 
the caloric curve of nuclei is found to follow very closely \cite{enterria:Melby99} the $\sim T^2$ Fermi law given 
by the well-known Bethe formula for the nuclear density of states \cite{enterria:Bohr98}.
Such a trend is maintained at moderately low nuclear excitation energies
($\epsilon^\star\lesssim$3{\it A} MeV) \cite{enterria:Wada89}.
In the region $\epsilon^\star\approx$ 3{\it A} - 8{\it A} MeV, however, the curve flattens 
\cite{enterria:Poch97} suggesting a phase transition (the width 
of the ``plateau'' indicating the latent heat related to the phase change). 
Although such thermodynamical behaviour has also been observed in other microscopic
systems such as metallic \cite{enterria:Schm01} and hydrogen \cite{enterria:Gobe01} clusters, 
the empirical determination (and the theoretical interpretation)
of the nuclear caloric curve has been a much debated issue the last years
\cite{enterria:Poch97,enterria:Gupt00,enterria:Rich00,enterria:Hirs99,enterria:Ma97,enterria:Nato01}.
As a matter of fact, the three experimental methods employed so far to measure the temperatures
attained in a reaction do not yield fully equivalent caloric curves.
The existing nuclear ``thermometers'' are based on: (i) the slopes of the kinetic energy spectra 
of light particles (n, p, $\alpha$) \cite{enterria:Morr94}, (ii) the 
double ratios of neighbour isotopes \cite{enterria:Albe89}, and (iii) 
the relative populations of excited states \cite{enterria:Serf98}.
Having in hand an alternative thermometer based on a clean and weakly
interacting probe would be extremely useful when searching for signals of the nuclear
liquid-gas phase transition. Electromagnetic probes, viz.
photons and dileptons, due to their weak final state interaction
with the surrounding medium, have long been recognized as the most
direct probes of the space-time evolution of the colliding nucleons
\cite{enterria:Cass90}. In HI reactions at intermediate bombarding energies, 
two concurrent processes are known to be responsible for 
hard-photon emission \cite{enterria:Nife85}: (i) first-chance (off-equilibrium)
collisions between projectile and target nucleons in the first
stages of the reaction, and (ii) subsequent $NN$ collisions in the produced
``nuclear fireball" zone. The experimental existence of a second
thermal component emitted from the produced hot nuclear sources and accounting for up 
to 30\% of the total photon yield above $E_\gamma$=30 MeV, has been unambiguously observed recently
\cite{enterria:Mart95,enterria:Schu97,enterria:Ente01,enterria:Orte02}. 
In this Letter we propose to exploit such radiation
as a novel thermometer of hot nuclear matter using a realistic thermal 
model which reproduces satisfactorily the observed photon spectral 
shapes and yields. Using such a thermometer we then construct the caloric 
curve in the region of the expected liquid-gas phase change.

\section{Thermal model}
Photons produced in HI reactions escape freely the interaction
region immediately after their production. Thus, even when emitted from an
equilibrated source they do not have a blackbody spectrum at the source 
temperature. However, the inverse slope parameter of their spectrum 
$E_0^t$ and the temperature $T$ of the nuclear medium are strongly 
correlated.  
In this work we employ the thermal model of Neuhauser and Koonin
\cite{enterria:Neuh87} (henceforth NK) to quantitatively relate $E_0^t$ 
and $T$. According to this model, the differential rate of photons emitted 
in incoherent $NN\gamma$ processes within a hot nuclear 
fragment is described by the expression:

\begin{equation}
\label{eq:koonin1}
\frac{d^5N_\gamma}{d^3x dt dE_\gamma}=
8\int\frac{d{\mathbf p}_{1i}}{(2\pi)^3}\frac{d{\mathbf p}_{2i}}{(2\pi)^3}
f({\mathbf p}_{1i})f({\mathbf p}_{2i}) \beta_{12i} \frac{d\tilde{\sigma}_\gamma}{dE_\gamma},
\end{equation}

where ${\mathbf p}_{1,2i}$ and $\beta_{12i}$ are the initial momenta
and relative-velocity of the colliding nucleons, $f({\mathbf p})$ their
(single-particle) momentum distribution, and $d\tilde{\sigma}_\gamma/dE_\gamma$ the
angle-integrated {\it Pauli-blocked} $NN$ Bremsstrahlung
cross-section. Approximating the emitting region as
nuclear matter in thermal equilibrium with local temperature $T$ 
and density $\rho$, the momentum distribution can be simply
parametrized by a hot Fermi-Dirac distribution ($\hbar=c=k_B=1$):

\begin{equation}
\label{eq:fermi-dirac}
f({\mathbf p})=\frac{1}{1+exp\left\{\left[\sqrt{p^2+m_N^2}-\mu(\rho)\right]/T\right\}},
\end{equation}

normalized to $4\int d{\mathbf p}f({\mathbf p})/(2\pi)^3 = \rho$. For low temperatures
($T\ll\epsilon_F$) the chemical potential $\mu$ of a system of nucleons can be written 
as a function of the Fermi energy $\epsilon_F(\rho)$ 
[where $\epsilon_F=\sqrt{p_F^2+m_N^2}-m_N$ and $p_F(\rho)=(3\pi^2\rho/2)^{1/3}$]
and $T$ \cite{enterria:Lope00}:

\begin{equation}
\label{eq:chemical potential}
\mu(\rho)\approx\epsilon_{F}(\rho)\left\{1-\frac{\pi^2}{12}\left[\frac{T}{\epsilon_{F}(\rho)}\right]^2\right\}.
\end{equation}

The {\it in-medium} Bremsstrahlung cross-section $d\tilde{\sigma}_\gamma/dE_\gamma$
of Eq. (\ref{eq:koonin1}) is approximately \cite{enterria:Neuh87}

\begin{equation}
\label{eq:koonin2}
\frac{d\tilde{\sigma}_\gamma}{dE_\gamma}\approx \frac{d\sigma_\gamma}{dE_\gamma}\int
\left[1-f({\mathbf p}_{1f})\right]\left[1-f({\mathbf p}_{2f})\right]\frac{d\Omega_\gamma}{4\pi}\frac{d\Omega_f}{4\pi},
\end{equation}

where $d\sigma_\gamma/dE_\gamma$ is the elementary $NN$
Bremsstrahlung cross-section in free space, $[1-f]$ the usual
Pauli-blocking factors, and $d\Omega_{\gamma,f}$ the solid angle of the outgoing
gamma and nucleons. At the considered energies, the reaction
$pp\rightarrow pp\gamma$ is suppressed by
a factor $\sim$20 with respect to $pn\rightarrow pn\gamma$
\cite{enterria:Scha91}, neutron-neutron Bremsstrahlung is vanishingly small, and 
thus one needs only to consider the $pn\gamma$
process. The isospin-averaged cross-section $d\sigma_\gamma/dE_\gamma$
is then one half of $d\sigma_{pn\gamma}/dE_\gamma$. We employ here 
the parametrization of Sch\"afer {\it et al.} 
for $d\sigma_{pn\gamma}/dE_\gamma$ derived within a proper relativistic 
and gauge-invariant meson-exchange effective model for the $NN$ 
interaction \cite{enterria:Scha91}, which reproduces well the available 
data on $pn$ Bremsstrahlung \cite{enterria:Male91}.

\subsection{Nuclear temperatures}
\begin{figure}[htbp]
\begin{center}
  \mbox{\epsfxsize=11.0cm \epsfbox{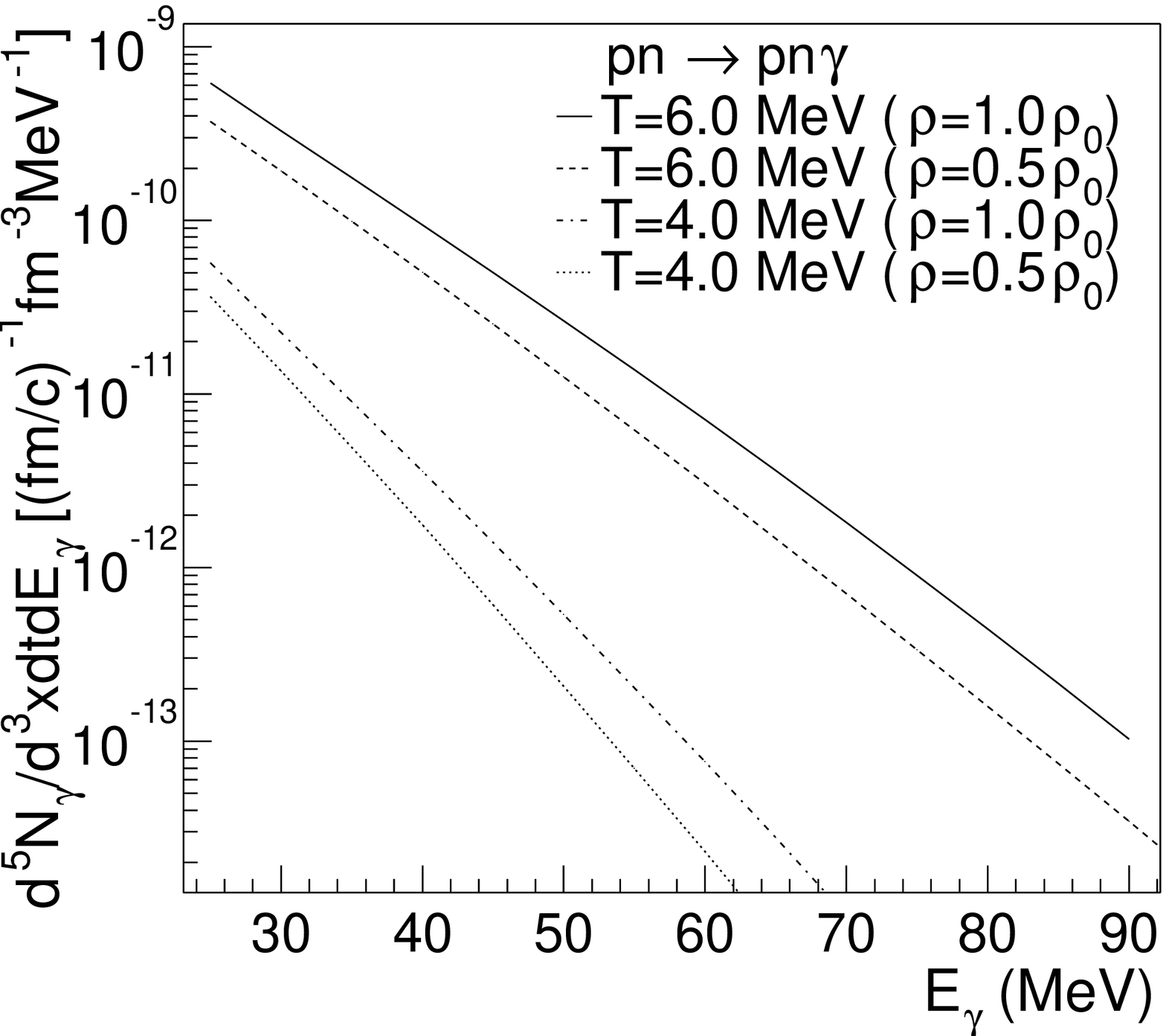}}
\end{center}
\caption{Thermal Bremsstrahlung emission rates, Eq.
(\protect\ref{eq:koonin1}), given by the NK model for a nuclear
system in equilibrium at temperatures $T$ = 4, 6 MeV and
densities $\rho/\rho_0$=0.5, 1.0.}
\label{fig:thermal koonin}
\end{figure}

\begin{figure}[htbp]
\begin{center}
  \mbox{\epsfxsize=11.0cm \epsfbox{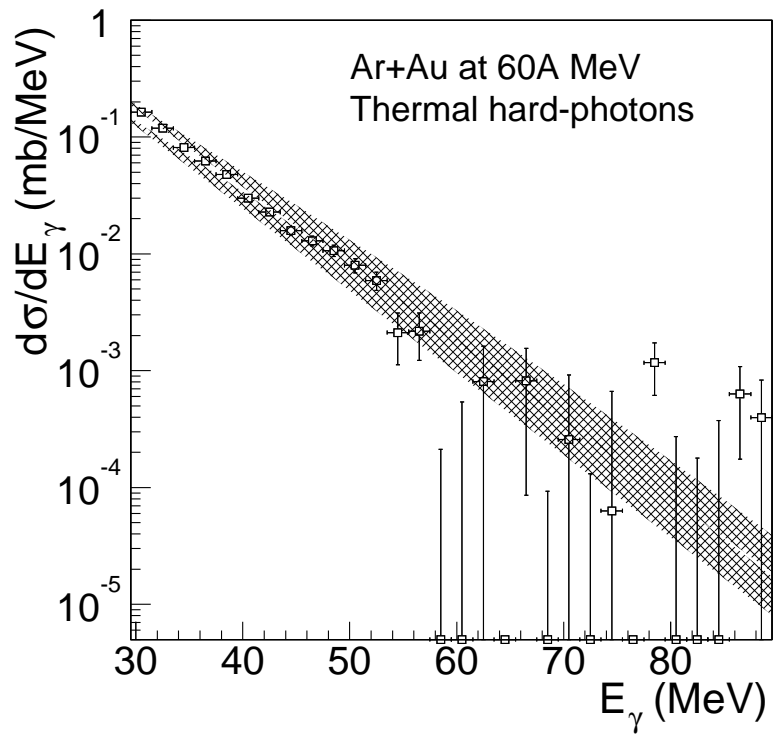}}
\end{center}
\caption{Thermal hard-photon spectrum measured in
the reaction $^{36}$Ar\/+\/$^{197}$Au at 60{\it A} MeV
\protect\cite{enterria:Ente01} compared (dashed band) to the prediction of
the NK model for a source at $\rho = \rho_0$ and $T$=5.3 $\pm$ 0.5 MeV (see text and Table
\protect\ref{tab:enterria:1}).}
\label{fig:thermal spec}
\end{figure}

The emission rates given by Eq. (1) can be thus calculated for a 
nuclear system at temperature $T$ and density $\rho$. 
The Bremsstrahlung rates are shown in Fig.
\ref{fig:thermal koonin} for a source at $T$ = 4, 6 MeV 
and $\rho/\rho_0$ = 0.5, 1.0 ($\rho_0$= 0.16 fm$^{-3}$). The
resulting hard-photon distributions above 30 MeV can be well
approximated by a Boltzmann exponential with slope $E_0^t$
in agreement with the experimental data (Fig. \ref{fig:thermal spec}). 
The integrated yields scale approximately with $T^{6.7}$ and $\rho$. 
The temperature $T$ of the emitting source and the photon slope 
parameter $E_{0}^{t}$, extracted from an exponential fit of the 
model spectra above $E_\gamma$ = 30 MeV, 
are found to be well described by the relation:

\begin{equation}
\label{eq:T vs E0t}
T(\mbox{MeV})\;=\;(0.78 \pm 0.02)\cdot E_0^t(\mbox{MeV}),
\end{equation}

in the range $T\approx$ 3 - 10 MeV and $\rho\approx (0.3 - 1.2)\rho_0$.

Applying Eq. (\ref{eq:T vs E0t}) the nuclear temperature attained
in a HI reaction can be determined by measuring the slope of
its thermal hard-photon spectrum. The temperatures obtained for all
the systems studied by the TAPS collaboration where a thermal
hard-photon component has been measured
\cite{enterria:Mart95,enterria:Schu97,enterria:Ente01,enterria:Orte02}
lie in the range $T\approx$ 4 - 7 MeV (Table \ref{tab:enterria:1})
in agreement with the typical values found in intermediate-energy
nucleus-nucleus collisions \cite{enterria:Gupt00,enterria:Rich00}.
The highest $T$ ($E_0^t$) values for a given incident energy $\epsilon_{lab}$
correspond to the most symmetric systems, which have the
highest energy deposition in the center-of-mass,
$\epsilon_{AA}=A_{red}\,\epsilon_{lab}/A_{tot}$, i.e., the largest
attainable excitation energies. In comparison with the nuclear thermometers used so far
\cite{enterria:Morr94,enterria:Albe89,enterria:Serf98}, the
thermal-photon thermometer presents several advantages: 
(i) minimal preequilibrium contamination due to the large difference, by
a factor two to three, between the direct (first-chance) and thermal hard-photon slopes
\cite{enterria:Schu97}; (ii) absence of final-state distortions like 
side feeding, rescattering, and reaccelaration by the Coulomb field and/or collective motion; 
(iii) measurement of the $T$ of the system right after equilibration, just before its break-up 
\cite{enterria:Ente01}, when the maximum thermal energy of the equilibrated residue is achieved; 
and (iv) intrinsic selection of (semi)central dissipative reactions 
with large number of $NN$ collisions \cite{enterria:Ente01}.
The use of thermal photon slopes as a nuclear thermometer is however
restricted to intermediate-energy reactions with $\epsilon_{lab}\approx$ 30{\it
A}-90{\it A} MeV. The lower limit in $\epsilon_{lab}$ results from 
the experimental difficulty to resolve accurately the thermal component in a double-exponential
fit of the hard-photon spectrum due to: (i) the very small cross-section in the
direct high-energy region\footnote{The emission probability of a photon of $E_\gamma>$ 30 MeV per $pn$
collision is a steeply increasing function of the incident energy
$\epsilon_{\mbox{\tiny{lab}}}$ \cite{enterria:Schu97}: $P_\gamma
\sim \mbox{exp}[-(1/\epsilon_{\mbox{\tiny{lab}}})]$.}, and (ii)
the increasing role at E$_{\gamma}<$ 20 MeV of statistical 
$\gamma$ from decays of Giant Dipole Resonances and bound states. 
Above $\epsilon_{lab}\approx$ 90{\it A} MeV, hard-photon spectra are 
well described by a single ``direct'' exponential \cite{enterria:Schu94,enterria:Mart99}
and also the background of photons from the decay 
of the produced $\pi^0$'s becomes important \cite{enterria:Mart99}.

\subsection{Photon multiplicities}
The absolute thermal photon yield per nuclear reaction,
$M_\gamma\equiv\sigma_\gamma/\sigma_{AA}$, predicted by the NK model can be obtained
integrating Eq. (\ref{eq:koonin1}) over the relevant space-time
history of the equilibrated system produced in a nucleus-nucleus
reaction:

\begin{equation}
\label{eq:gamma multiplicity}
M_\gamma^{\mbox{\tiny{NK}}}=
\int d^3x\int dt\int^{\infty}_{E_\gamma=30\mbox{ \tiny{MeV}}}
dE_\gamma\frac{d^5N_\gamma}{d^3x dt dE_\gamma}(T,\rho).
\end{equation}

In the most general case $T=T(x,t)$ and $\rho=\rho(x,t)$, and the
integration in Eq. (\ref{eq:gamma multiplicity}) requires a consistent
modeling of the space-time evolution of the reaction as done, e.g.,
in hydrodynamical approaches. Since we are
just interested in verifying that the thermal model provides also a correct 
estimation of the measured thermal photon yields, we will significantly simplify
the calculation of $M_\gamma^{\mbox{\tiny{NK}}}$ making a few plausible assumptions.
First, one may neglect any $T$ and $\rho$ gradients within the nuclear
source and approximate the integral over space by the volume of the participant nucleons 
in the reaction, $V=A_{tot}/\rho$, where $A_{tot}= A_1+A_2$ is the sum of target 
and projectile nucleons\footnote{The equilibrated nuclear object produced in (semi)central 
collisions at these energies is likely to include many more constituents than just the original 
participants as the excitation is shared with other nucleons. Although it is improbable that 
{\it all} nucleons are available to participate in the nuclear Bremsstrahlung process 
(e.g. 10\%-20\% of them may just be gone in preequilibrium emission), we take here 
$A_{tot}= A_1+A_2$ as a reliable measure of the (relative) size of the system in the different reactions.}. 
Secondly, 
one can simply replace the integral over time by a constant interval equal to the 
lifetime of the radiating source, $\Delta\tau\approx$ 100 fm/c, consistent with the 
usual estimates of the lifetime of equilibrated excited nuclei \cite{enterria:Rich00}. 
With these rough approximations, partial integration of 
Eq. (\ref{eq:koonin1}) above $E_\gamma$= 30 MeV permits to compute
$M_\gamma^{\mbox{\tiny{NK}}}$ straightforwardly. 
Since the emission rates given by Eq. (1) scale as $\sim\rho$ 
and since $V\propto\rho^{-1}$, the thermal $\gamma$ multiplicities are rather 
insensitive to the density of the source in this very simple estimation.
Most experimental multiplicities (Table \ref{tab:enterria:1}) are very
well reproduced by the integrated NK photon rates and our simplified ansatz 
of the space-time history of the radiating source\footnote{The reported errors on 
$M_\gamma^{\mbox{\tiny{NK}}}$ in Table \ref{tab:enterria:1} include the uncertainties in $T$ 
propagated from $E_0^t$ and are quite large due to the strong power dependence of 
the photon emission rates on $T$.}. The photon yields of the two systems with largest
excitation energies ($^{86}$Kr+$^{58}$Ni and $^{36}$Ar+$^{58}$Ni at 60{\it A} MeV), however, are
overpredicted by the model (though still within the combined $M_{\gamma}^{\mbox{\tiny{NK}}}$
and $M_{\gamma}^{\mbox{\tiny{exp}}}$ systematic uncertainties). A much better agreement would be
reached in these two cases if the lifetime of the equilibrated sources was reduced by a factor $\sim$2 
compared to lower energies (i.e. if $\Delta\tau\sim$ 50 fm/c). This result is in quantitative 
agreement with the observed fast break-up of nuclear systems for increasingly larger values of
$\epsilon^\star$ interpreted in terms of a low-density (spinodal) fragmentation \cite{enterria:Beau00}.

\section{Caloric curve}
\begin{figure}[htbp]
\begin{center}
  \mbox{\epsfxsize=11.0cm \epsfbox{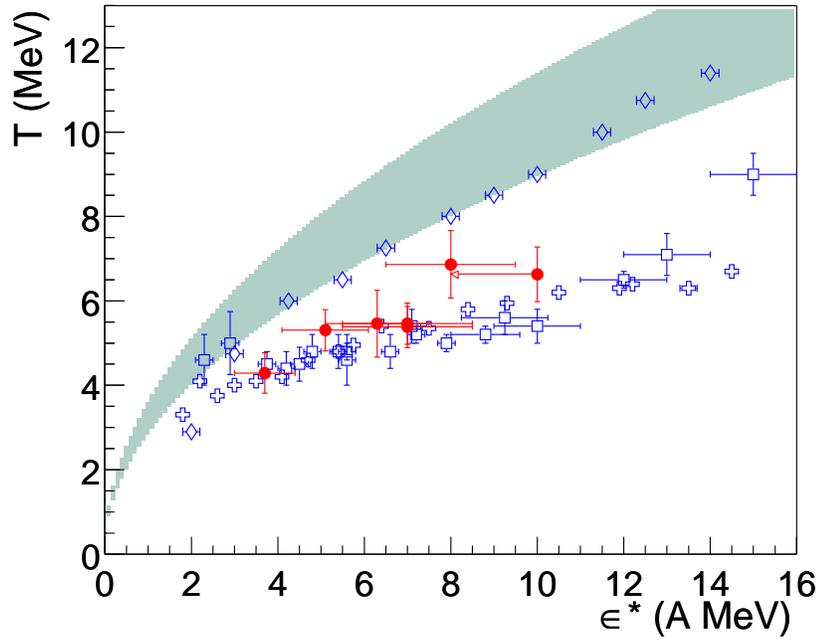}}
\end{center}
\caption{Caloric curve constructed with the photon slope thermometer (dots)
compared to ALADIN (squares) and EOS (crosses) curves (isotopic $T$'s),
and to INDRA (rhombbi) curve (kinetic $T$'s). 
The dark area corresponds to the Fermi relation $\epsilon^\star=aT^2$ with $a=A/8-A/13$ MeV$^{-1}$.}
\label{fig:new caloric curve}
\end{figure}

Having determined the values of $T$ from $E_0^t$ through Eq. (\ref{eq:T vs E0t}) 
for our different reactions, a caloric curve $\epsilon^\star(T)$, can be constructed
correlating $T$ with the {\it thermal} excitation energies $\epsilon^\star$ attained 
in each reaction. To construct such a curve, we use the published values of
$\epsilon^\star$ measured in semi-central and central reactions with
equivalent systems (Table \ref{tab:enterria:1}, most right column). 
In the case of $^{86}$Kr+$^{58}$Ni at 60{\it A}, 
in the absence of an experimental measurement of $\epsilon^\star$, an upper value can be
obtained from the total (Coulomb-corrected) center-of-mass energy,
$\epsilon_{AA}$, 
subtracted of other energy contributions\footnote{$\epsilon^\star\approx(\epsilon_{\mbox{\tiny{AA}}}+Q)-\epsilon_{\mbox{\tiny{coll}}}-\epsilon_{\mbox{\tiny{preeq}}}-\epsilon_{\mbox{\tiny{rot}}}$, where: (i) the reaction Q-value and $\epsilon_{\mbox{\tiny{rot}}}$ are of the
order $\sim$1{\it A} MeV and have opposite signs, and (ii) the
radial flow energy, $\epsilon_{\mbox{\tiny{coll}}}$, and the pre-equilibrium
component, $\epsilon_{\mbox{\tiny{preeq}}}$, represent $\sim$3{\it A} MeV for a
symmetric system at these energies \cite{enterria:Poch97,enterria:Gupt00,enterria:Rich00}.}.
Although it is fair to acknowledge that our globally indirect determination of 
$\epsilon^\star$ is potentially subject to significant systematic uncertainties, 
such uncertainties are reasonably contained in the already intrinsically large 
experimental errors of the excitation energy measurements 
\cite{enterria:Gupt00} and, 
furthermore, the qualitative conclusions that we obtain from the analysis of the photon
caloric curve would only be modified by much larger shifts along the
$\epsilon^\star$ axis. The $\epsilon^\star$ - $T$ correlation for the 7 reactions
considered here is shown as solid dots in Fig. \ref{fig:new caloric
curve}. 
Our caloric curve falls somewhat above the ALADIN \cite{enterria:Poch97} 
and EOS \cite{enterria:Haug96} results\footnote{Note however that both, ALADIN and EOS, 
collaborations have updated slightly upwards their original curves 
(shown in Fig. \ref{fig:new caloric curve}) to account for several corrections in $T$ (sequential feedings) and 
$\epsilon^\star$ (collective radial flow energy), see e.g. \cite{enterria:Nato01} for a 
compilation of the most recent results.} obtained with two different isotopic ratios,
but still clearly below the region expected for a pure degenerate Fermi fluid
(Fig. \ref{fig:new caloric curve}, dark band). The fact that the thermal photon
temperatures seem to be slightly larger than the isotopic ones 
is not surprising since ``chemical'' temperatures probe a later lower
density (cooler) stage when isotopes are produced, whereas the most significant part of
thermal photon emission takes place somewhat earlier. Indeed in this scenario, the 
radiation of thermal Bremsstrahlung photons occurs during the path followed by 
the excited nuclear system between the hot ``liquid'' phase (one single big nuclear 
fragment at the end of the pure pre-equilibrium stage) 
and the mixed phase of gaseous particles and small droplets produced later in 
the fragmentation process \cite{enterria:Ente01}.
Our caloric curve disagrees, on the other hand, with the one obtained by the 
INDRA collaboration using kinetic temperatures \cite{enterria:Ma97}, 
which follows closely the $\epsilon^\star\sim T^2$ trend.
The kinetic $T$'s are systematically higher by about 1-2 MeV 
most likely due to the fact that the slope parameters
of $p$ and $\alpha$ kinetic energy spectra are influenced by
preequilibrium and mid-rapidity emissions \cite{enterria:Dore00}, and/or nucleon Fermi motion effects.
The departure of the photon caloric curve from the 
$\epsilon^\star=aT^2$ law (with level density parameters $a=A/8-A/13$ MeV$^{-1}$) 
characteristic of the nuclear ground state (liquid phase), and the 
slow increase of $T$ with $\epsilon^\star$ in the region of intermediate excitation
energies are in qualitative agreement with the expected behaviour of a 
liquid-gas phase transition occurring in excited atomic nuclei.

\section{Conclusions}
A thermal bremsstrahlung model has been employed to extract the 
thermodynamical properties of the nuclear systems produced in heavy-ion 
collisions at intermediate-energies. Such a model predicts exponential
hard-photon spectra above $E_\gamma$ = 30 MeV in
agreement with the experimental data. The thermal slopes are
linearly correlated with the temperature of the emitting system.
This defines a new thermometer of hot nuclear matter which, at
variance with the usual hadronic-based methods, is free
of significant final-state distortions. 
The hot nuclear residues prepared in different reactions with
excitation energies $\epsilon^\star \approx$ 3{\it A} - 8{\it A}
MeV, have temperatures in the range $T\approx$ 4 - 7 MeV yielding
a slowly increasing caloric curve which lies below the expected behaviour 
for a pure Fermi liquid.

\section{Acknowledgements}

We thank the members of the TAPS collaboration who participated in
the KVI and GANIL experimental campaigns in 1997 and 1998. This
work has been in part supported by an IN2P3-CICYT agreement, by
the Dutch Foundation FOM, and by the European Union HCM network 
under Contract No. HRXCT94066. D.G. d'E. has been supported by the European Union
TMR Programme (Marie-Curie Fellowship No. HPMF-CT-1999-00311).





\newpage

\clearpage

\begin{table}
\begin{center}
\caption{Heavy-ion reactions studied by the TAPS 
collaboration where a thermal bremsstrahlung component has been
identified. For each reaction we report: (i) the nuclear
temperatures $T$ extracted through Eq. (\protect\ref{eq:T vs E0t})
from the measured thermal slopes $E_0^{t}$; (ii) the total number of
nucleons $A_{tot}$, the photon multiplicity
$M^{\mbox{\tiny{NK}}}_{\gamma}$ predicted by the NK model (for a
source with volume $V=A_{tot}/\rho$, and lifetime $\Delta\tau$=100 fm/c) 
compared to the experimental value $M^{\mbox{\tiny{exp}}}_{\gamma}$; 
and (iii) the excitation energies $\epsilon^\star$ measured in other 
experiments.}
\resizebox{16.5cm}{!}{ 
\begin{tabular}{lc|cc|ccc|c}
\hline
System  & $\epsilon_{\mbox{\tiny{lab}}}$ ({\it A}MeV) &  $E_0^{t}$ (MeV) & $T$ (MeV) & $A_{\mbox{\tiny{tot}}}$ & $M^{\mbox{\tiny{NK}}}_{\gamma}$ (10$^{-4}$)& $M^{\mbox{\tiny{exp}}}_{\gamma}$ (10$^{-4}$) & $\epsilon^\star$ ({\it A}MeV)\\\hline
$^{208}$Pb+$^{197}$Au & 30 & 5.5 $\pm$ 0.6 \cite{enterria:Mart95} & 4.3 $\pm$ 0.5 & 405 & 0.7 $\pm$ 0.7 & 0.6 $\pm$ 0.2 \cite{enterria:Mart95} & 3.7 $\pm$ 0.7 \cite{enterria:Leco94}\\
$^{36}$Ar+$^{197}$Au  & 60 & 6.8 $\pm$ 0.6 \cite{enterria:Ente01} & 5.3 $\pm$ 0.5 & 233 & 1.9 $\pm$ 1.3 & 1.6 $\pm$ 0.2 \cite{enterria:Ente01} & 5.1 $\pm$ 1.0 \cite{enterria:Sun00}\\
$^{181}$Ta+$^{197}$Au & 40 & 6.9 $\pm$ 0.6 \cite{enterria:Mart95} & 5.4 $\pm$ 0.5 & 378 & 3.4 $\pm$ 1.9 & 3.2 $\pm$ 1.0 \cite{enterria:Mart95} & 7.0 $\pm$ 1.5 \cite{enterria:Norm00}\\
$^{36}$Ar+$^{107}$Ag  & 60 & 7.0 $\pm$ 1.0 \cite{enterria:Ente01} & 5.5 $\pm$ 0.8 & 143 & 1.5 $\pm$ 1.5 & 1.2 $\pm$ 0.2 \cite{enterria:Ente01} & 6.3 $\pm$ 1.2 \cite{enterria:Sun00}\\
$^{129}$Xe+$^{112}$Sn & 50 & 7.0$\pm$ 0.6 \cite{enterria:Orte02}  & 5.5 $\pm$ 0.6 & 241 & 3.1 $\pm$ 2.2 & 2.6 $\pm$ 0.4 \cite{enterria:Orte02} & 7.0 $\pm$ 1.5 \cite{enterria:Boug99}\\
$^{86}$Kr+$^{58}$Ni   & 60 & 8.5 $\pm$ 0.8 \cite{enterria:Mart95} & 6.6 $\pm$ 0.6 & 144 & 4.4 $\pm$ 2.7 & 2.0 $\pm$ 0.4 \cite{enterria:Mart95} & $<$10. (see text)\\
$^{36}$Ar+$^{58}$Ni   & 60 & 8.8 $\pm$ 1.0 \cite{enterria:Ente01} & 6.9 $\pm$ 0.8 &  94 & 4.1 $\pm$ 2.9 & 1.1 $\pm$ 0.2 \cite{enterria:Ente01} & 8.0 $\pm$ 1.5 \cite{enterria:Sun00}\\
\hline
\end{tabular}
}
\label{tab:enterria:1}
\end{center}
\end{table}

\end{document}